\documentclass[manuscript,screen]{acmart}

\AtBeginDocument{%
  \providecommand\BibTeX{{%
    \normalfont B\kern-0.5em{\scshape i\kern-0.25em b}\kern-0.8em\TeX}}}

\begin{document}

\title{The privacy protection effectiveness of the video conference platforms' virtual background and the privacy concerns from the end-users}

\author{Shijing He}
\email{saiking.edward@gmail.com}
\affiliation{%
  \institution{Cisco Systems, Inc.}
  \city{Shanghai City}
  \country{China}
}

\author{Yaxiong Lei}
\affiliation{%
  \institution{University of St.Andrews}
  \city{St.Andrews}
  \country{UK}}
\email{yl212@st-andrews.ac.uk}

\renewcommand{\shortauthors}{S.He and Y.Lei}

\begin{abstract}
Due to the abrupt arise of pandemic worldwide, the video conferencing platforms are becoming ubiquitously available and being embedded into either various digital devices or the collaborative daily work. Even though the service provider has designed many security functions to protect individual's privacy, such as virtual background (VB), it still remains to be explored that how the instability of VB leaks users' privacy or impacts their mentality and behaviours. In order to understand and locate implications for the contextual of the end-users' privacy awareness and its mental model, we will conduct survey and interviews for users as the first stage research. We will raise conceptual challenges in terms of the designing safety and stable VB, as well as provide design suggestions.
\end{abstract}



\maketitle

\section{Introduction}
Due to the global circumstances caused by the COVID-19 pandemic, an unprecedented increase in working from home (WFH) or remotely working happens worldwide. Everything has become "remotely", and WFH is getting a facelift and becoming more popular. Video conference platforms (e.g. Zoom, Webex and Microsoft Teams) have played an alternative role, and have replaced the office-based work tradition. Nowadays, people are getting used to the WFH mode. Bloom et al. ~\cite{bloom2015does} noticed WFH could increase productivity by 13\%. The reason why the performance increase lies in the quiet and convenient working environment that could lead to more calls per minute.

In addition, the aforementioned platforms and companies continue to actively optimize global customer's feedback and usage worldwide, and provide several new functions to protect the meeting security. Virtual background (VB), as one of a privacy protection design, has been widely applied on different video conference platforms. Users could apply VB to limit distractions and maintain the surrounding environment to avoid the privacy leaking, as well as satisfy their aesthetic needs. VB also allows users to blur background, change pictures or customize background accordingly. In particular, Yao et al.~\cite{yao2017privacy} has pointed out that blurring is an effective method to protect people's sensitive data. 

Although the VB has provided a good chance to show user's personality (aesthetic taste or personal preference), or as a business requirement (sales meet with customers on behalf of the company), education demands (academic conference, teaching needs), people started to concern about whether the VB will leak their privacy or not. The instability of the segmentation algorithm also has its limitations on leaking people’s environment data or potential privacy. Our future research will aim to investigate this question and explore the privacy mental model of participants, as well as their privacy awareness in different scenarios.

\begin{figure}
    \centering
    \includegraphics[width=\textwidth]{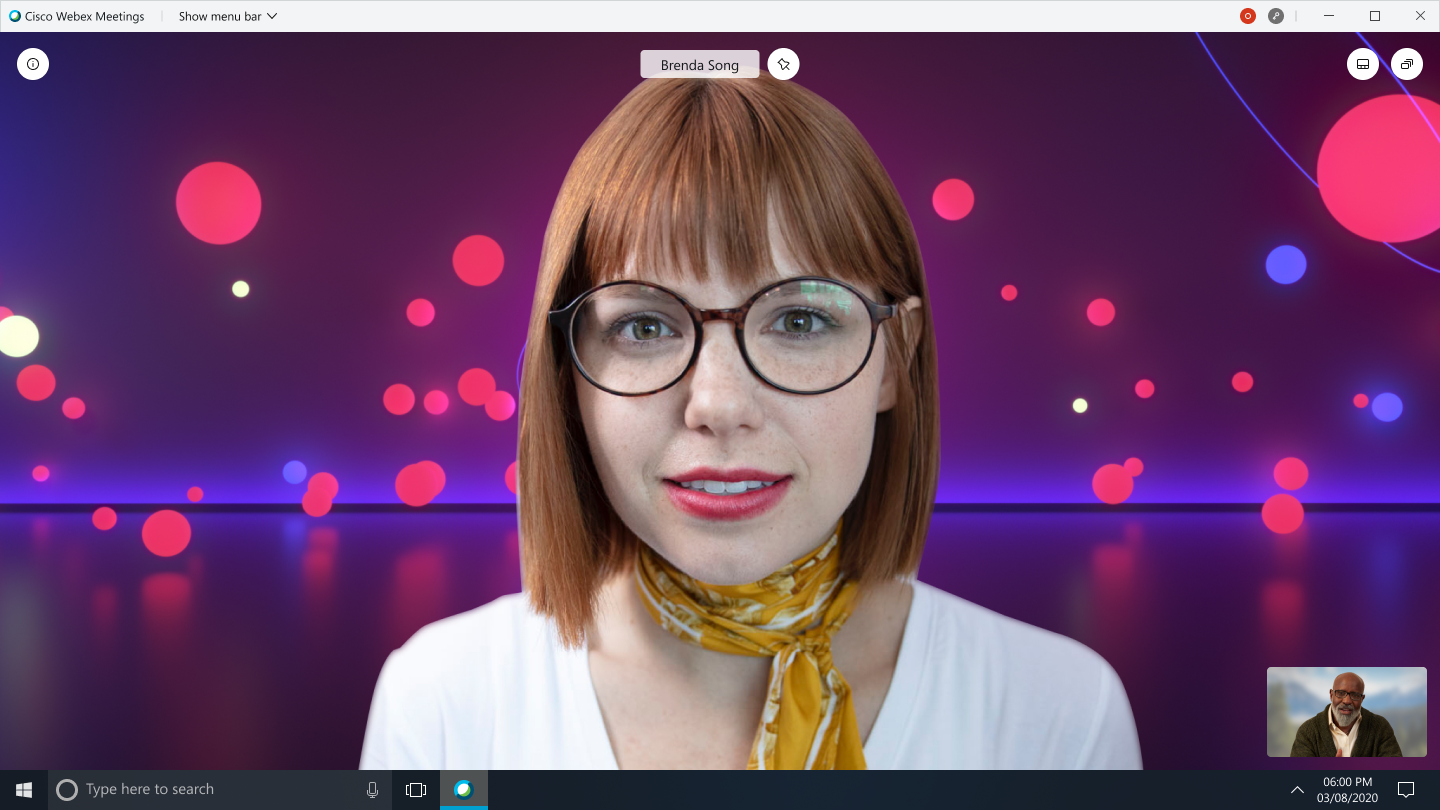}
    \caption{2020 Cisco. Webex Virtual Background. (\url{https://blog.webex.com/wp-content/uploads/2020/08/Virtual-Background-in-Meeting-Spheres-e1596570058514.png})}
    \label{fig:my_label}
\end{figure}

\section{Related works}
There has been a wide investigation about using video conference platforms as effective remotely virtual tools in different fields. For instance, in educational field ~\cite{barbosa2019zoom,serhan2020transitioning,lowenthal2020thinking,kohnke2020facilitating},qualitative data collection and research method ~\cite{gray2020expanding,archibald2019using}, Bailenson ~\cite{bailenson2021nonverbal} has provided 4 explanations to furtherly illustrate the causes of the Zoom fatigue: "excessive amounts of close-up eye gaze, cognitive load, increased self-evaluation from staring at video of oneself, and constraints on physical mobility." Due to the amount of time in video conferencing, nearly half users prefer to work from the living room or bedroom ~\cite{Nulab20}. Conti et al. ~\cite{conti2021not} also noticed the similar privacy concerns about VB, that is, users could conceal where they are with the VB for the purpose of cheating other participants in the meeting.

Online meeting or live streaming platforms, similar to aforementioned video conferencing platforms, have provided VB features and offered some room for customization. For example, users can upload their own preferred VB images. By semantic segmentation of each frame in the video in real time, the algorithm separates the user from the surrounding environment and replaces it with a VB. Video meeting as a high-frequency scenario in daily work, along with the popularity of WFH, the protection of user privacy has put forward a higher demand.

Compared with the high-performance servers in the cloud, the video conferencing carriers of personal scenarios are mainly laptops of various colors. There are three main factors---various models of laptops’ performances,  highly real-time requirements of video conferencing, various online-meeting contexts,  that have put more demanding demands on the performance of the end-side (or say user-side) algorithm. 

The implementation of the real-time feature needs the end-side portrait segmentation model to be light enough, while such a small model processing requires more sensitive data and is weak in some difficult scenarios (such as portrait edge and background similarity, etc.). These will easily lead to background misclassification into portraits, blurred portrait edges, etc \cite{lin2021real}. These problems have directly resulted in the performance and accuracy of the algorithm, the drawback of the segmentation algorithm would cause the virtual background to be unstable. There is also a certain privacy exposure problem, along with the aliasing and other segmentation problems, which will lead to the risk of potential leakage to the environment behind the participants.  

\section{Research Aim and Question}
According to the internal user experience feedback and user testing data when using Cisco Webex VB from Webex Design Team, we have found that the circumstance of WFH could bring a certain risk of an individual's privacy leakage, especially if users apply the VB or blur background. We will not only study how privacy concerns proceed practically in a range of circumstances or scenarios with classic qualitative research approaches, but also be grounded in actual design practices. We believe that the experimental result will be more likely to be adopted. Our main research questions are:  

\textbf{RQ1:} What is the possibility of causing  panic and discomfort for end-users when there are sudden privacy leaks caused by VBs in meetings?

\textbf{RO2:} Is the power differential ~\cite{bernd2020bystanders} or identity (e.g., interviewer and interviewee) in meetings as a key factor that can impact the mental model of privacy awareness when the sudden privacy leaks caused by VBs in meetings?

To answer these questions, we will conduct an empirical investigation from the end-user's perspective, which contains scenario-based surveys (see Methodology and Experimental Design).  We will conduct semi-structured interviews based on survey results, in order to provide personal information and insights in the context of using video conferencing platforms.

\section{Methodology and Experimental Design}
The early stage of research would be based on the Grounded theory that was applied to the existing privacy qualitative studies ~\cite{abdi2021privacy,westin2021s}. In particular, the interviewing and analysis will be conducted simultaneously and iteratively by using the Grounded theory. We will conduct online surveys for end-users, and set semi-structured interviews for the respondents who joined the survey. This aims to investigate the privacy perception and personal awareness in different scenarios and contextual environment factors. The participants will be the experienced end-users. Moreover, the bystanders (The bystander types are basically include the cohabitants, acquaintances or strangers (e.g. delivery man or someone suddenly flashed ), or other potential factors maybe impact the meeting status, such as pets, or sudden accidents.) will also be evaluated because they may appear in the camera randomly, due to the instability of the recognition algorithm. This might be a good start to understand the end-user's mental model in which the respondents face the emergency situation in the various scenarios; as well as to investigate the differences of the power differentials and identity. To sum up, we will focus on end-users and their mental model depending on contexts and scenarios. 

Additionally, including aforementioned cases, we also have  extended  the privacy factors and dimensions that containing: emergencies (e.g. external influence, technical limitation), the number of participants, social relations (stranger-intimacy), power differential, meeting attributes (official or unofficial), benefit, network types, duration time, recurrence, environment and number of bystanders. Considering different privacy awareness factors, we will create six main scenarios---- 1) Business requirements, 2) Education demands, 3) Family meeting, 4) Job interview, 5) Online conference or event and  6) Enterprise internal meeting.

The ethics forms, survey and remuneration mechanism will be reviewed and provided by the Institutional Review Board (IRB) of the universities. The researchers who also participate in the related training will  get the certificates. Once complete the pilot test and final version data collection from Qualtrics, we will conduct interviews for participants who joined the survey investigation. The series of surveys consisted of multiple types of questions in both open-ended and closed-ended form.

\section{Future Work}
The future work will basically conduct in two mainly direction based contextual scenarios: the mental model of privacy awareness and algorithm optimization. To meet up with the expected result, we will continue to evaluate the mental model and user's privacy awareness so as to face future emergency situations. However, it is a bit difficult to observe the participants' behaviours directly and immersively as the lab-based interview could cause bias due to the Hawthorne effect. Hence, we will apply an ethnographic approach~\cite{gan2020connecting} and use quantitative data analysis (e.g., mental workload data or gaze data) to conduct a field study to get participants’ behaviours and micro-expression data from closer observation. In order to improve the existing user experience and the effectiveness of privacy protection, we will expect to deliver the end-users' mental model of privacy awareness. Moreover, due to the special contextual integrity (e.g., environment) of WFH, it should also be noticed whether these WFH circumstances are affected by personal smart home devices. For instance, the smart home device could potentially track voice data or even sensitive data when a user is having a confidential meeting. 

In addition, the algorithm design strategy has a noticeable impact on the VB's privacy improvement, improving the segmentation quality and effectiveness of algorithms is also urgently needed. Nowadays, due to the processing performance of end devices, lightweight models are usually used for portrait segmentation. Some work \cite{lin2021real} applied the idea of PointRend \cite{kirillov2020pointrend} to optimize the edge blurring problem and the performance problem under high resolution image conditions. It has to be noticed that only a few areas need fine segmentation and most areas need coarse segmentation. The problem of information exposed from coarse segmentation should also be brought to our attention.

\bibliographystyle{ACM-Reference-Format}
\bibliography{Reference}

\end{document}